\documentclass[10pt,showpacs,preprintnumbers,amsmath,eqsecnum,
amssymb,twocolumn,aps,prd,nofootinbib,
a4paper,superscriptaddress]{revtex4}
\pdfoutput=1

\usepackage{graphicx}

\def\be{\begin{equation}}
\def\ee{\end{equation}}

\def\bea{\begin{eqnarray}}
\def\eea{\end{eqnarray}}

\def\vp{{\varphi}}

\def\cs2{c_{\rm{s}}^2}

\def\U0{{\bar U_0}}

\def\bi{\begin{itemize}}
\def\ei{\end{itemize}}

\newcommand\eq[1]{Eq.~(\ref{#1})}

\def\Mpc{{\rm{Mpc}}}
\newcommand{\Mpl}{\ensuremath{M_\mathrm{PL}}}

\usepackage{epsfig}
\usepackage{graphicx,epsf}
\usepackage{color}
\usepackage{bm}
\usepackage{psfrag}

\def\be{\begin{equation}}
\def\ee{\end{equation}}
\def\beb{\begin{equation*}}
\def\eeb{\end{equation*}}
\def\bea{\begin{eqnarray}}
\def\eea{\end{eqnarray}}
\def\beab{\begin{eqnarray*}}
\def\eeab{\end{eqnarray*}}
\def\nn{\nonumber}

\def\vp{{{\varphi}}}

\def\w{{\omega}}
\def\cs2{c_{\rm{s}}^2}

\def \beg {\begin{enumerate}}
\def \en {\end{enumerate}}

\def\cs{c_{\rm{s}}^2}

\begin{document}
\title{Calculating Non-adiabatic Pressure Perturbations during Multi-field Inflation}
\author{Ian~Huston}
\email[]{i.huston@qmul.ac.uk}
\affiliation{Astronomy Unit, School of Physics and Astronomy, Queen Mary University of London,
Mile End Road, London, E1 4NS, UK}

\author{Adam J.~Christopherson}
\email[]{Adam.Christopherson@nottingham.ac.uk}
\affiliation{School of Physics and Astronomy, University of Nottingham, University Park,
Nottingham, NG7 2RD, UK}

\date{\today}

\begin{abstract}
Isocurvature perturbations naturally occur in models of inflation consisting of more 
than one scalar field. In this paper we calculate the spectrum of isocurvature perturbations
generated at the end of inflation for three different inflationary 
models consisting of two canonical scalar fields. The amount of non-adiabatic pressure present at
the end of inflation can have observational consequences through the generation of vorticity and
subsequently the sourcing of B-mode polarisation. We compare two different
definitions of isocurvature perturbations and show how these quantities evolve in different ways
during inflation. 
Our results are calculated using the open
source Pyflation numerical package which is available to download. 
\end{abstract}

\pacs{98.80.Cq \hfill arXiv:1111.6919}

\maketitle

\section{Introduction}
\label{sec:intro}

Inflationary models consisting of more than one scalar field have been long known to
create isocurvature, or non-adiabatic pressure, perturbations due to the relative
entropy perturbation between the different fields \cite{Kofman:1985zx, Linde:1985yf}. Isocurvature perturbations
during inflation are interesting since, for example, they induce the evolution of the
curvature perturbation on super-horizon scales \cite{GarciaBellido:1995qq, Rigopoulos:2003ak, Wands2000},
and can give rise to large non-gaussianities \cite{Langlois:2008vk, Kawasaki:2008sn}.

The fraction of isocurvature perturbations present around recombination is derived from observations of the 
 cosmic microwave background (CMB) such as those made by the {\sc Wmap} satellite,
and there have been several recent works studying 
this \cite{Bean:2006qz, GarciaBellido:2004bb, Bucher:2004an, Parkinson:2004yx}.
However, even in the `era of precision cosmology'  the {\sc Wmap} seven year results allow for some isocurvature, 
constraining the power spectrum of entropy perturbations to be of order 10\% that of the usual adiabatic perturbations
\cite{WMAP7}.

The most common method for computing perturbations from inflation is to use the
so-called `$\delta N$ formalism' \cite{Sasaki:1995aw, Lyth:2004gb, Lyth:2001nq}
which is a gradient expansion relating the perturbed number of e-folds to the curvature perturbation
on uniform density hypersurfaces. This is a concise method, but has some draw-backs, primarily that
it is only valid on super-horizon scales. 

An alternative method for computing inflationary perturbations is to use the full 
cosmological perturbation theory, a powerful 
technique which has been honed during the past few decades \cite{Bardeen:1980kt, ks, mfb} (see also
Refs.~\cite{MW2008, Malik:2008yp} and references therein for a comprehensive list). It
is this approach which we adopt in this article. This will enable us to consider the non-adiabatic pressure 
perturbation during general multi-field inflation and obtain its power spectrum at the end of inflation.

There are several reasons for wanting to consider the non-adiabatic pressure perturbation 
produced towards the end of inflation
in such a comprehensive manner. One interesting reason is the recent realisation that a non-adiabatic pressure
perturbation can source vorticity at second order in cosmological
perturbation theory \cite{vorticity, Christopherson:2010dw, Christopherson:2010ek}.
This vorticity will, in turn, have 
an effect on the CMB, since vector perturbations naturally source
B-mode, or rotational,
polarisation of the CMB radiation \cite{Hu:1997hv}. They can also possibly source primordial
magnetic fields \cite{Fenu:2010kh}
thus providing another way in which to distinguish between the many different inflationary models 
and help to constrain the inflationary paradigm. 

In this article we will consider several models of inflation consisting of two canonical scalar
fields which produce a non-negligible non-adiabatic signal,
and calculate the spectrum of entropy perturbations for these models in two different ways. 
The first quantity is the non-adiabatic pressure perturbation defined in terms of the total fluid
perturbations of pressure and energy density and is denoted $\delta P_\mathrm{nad}$. This is
compared with the spectrum of the rotated field values in the isocurvature direction $\delta s$.
We consider our results in comparison with those previously obtained and comment on the how well the
models we investigate are constrained by current observational data. We also show that the full
non-adiabatic pressure perturbation, $\delta P_\mathrm{nad}$, evolves in a quite
different way to $\delta s$. It is important to
distinguish between these quantities in evaluating the potential of particular physical models when
considering the generation of vorticity. The numerical code used in these calculations has been
released under an open source license and is available to download \cite{pyflation}.

This paper is organised as follows: in the next section we derive expressions for the non-adiabatic pressure perturbations
in multi-field inflation. We then outline our numerical method in Section~\ref{sec:numerical} and
present our results in Section~\ref{sec:results}. We discuss our conclusions and directions
of possible future work in Section~\ref{sec:discussion}.

\section{Preliminaries}

An adiabatic system is one in which 
\be 
\frac{\delta X}{\dot{X}} = \frac{\delta Y}{\dot{Y}}\,,
\ee 
for any scalars $X$ and $Y$, and where an overdot denotes the derivative with respect to coordinate time, $t$.
 It is usual to describe a fluid using its pressure and energy density,
$P$ and $\rho$, and so an adiabatic fluid is defined as one in which
\be 
\frac{\delta P}{\dot{P}} = \frac{\delta \rho}{\dot{\rho}}\,.
\ee 
In general, though, a fluid system is not adiabatic, and so the
pressure perturbation can be expanded as 
\be 
\label{eq:dPsplit}
\delta P = \delta P_{\rm nad} + c_{\rm s}^2\delta\rho\,,
\ee
where $\delta P_{\rm nad}$ is the non-adiabatic pressure perturbation and $c_\mathrm{s}^2 =
\dot{P}/\dot{\rho}$ is the adiabatic sound speed for the fluid. 

For a system consisting of more than
one fluid or field, this can then be expanded further into an intrinsic and relative entropy
perturbation \cite{ks}
\be 
\delta P_{\rm nad} \equiv \delta P_{\rm intr} + \delta P_{\rm rel}\,.
\ee

In this article we do not consider this split, and instead we use Eq.~(\ref{eq:dPsplit}) to 
calculate the value of $\delta P_\mathrm{nad}$ as
\be 
\delta P_{\rm nad}\equiv \delta P - c_{\rm s}^2\delta\rho\,.
\ee

In order to calculate the pressure and density of the system we consider linear, scalar perturbations 
to a flat Friedmann-Robertson-Walker spacetime, and choose the gauge in which
spatial hypersurfaces remain unperturbed -- the uniform curvature gauge. The line element thus takes the form
\be 
ds^2=-(1+2\phi)dt^2+2 a(t) B_{,i} dt dx^i+a^2(t)\delta_{ij}dx^i dx^j\,,
\ee 
where $a(t)$ is the scale factor, $\phi$ is the lapse function and $B$ is the scalar shear. We
consider a system of two canonical scalar fields with the 
Lagrangian
\be 
{\mathcal L}=\frac{1}{2}\Big({\dot{\vp}^2}+\dot{\chi}^2\Big)+V(\vp,\chi)\,.
\ee
%


The total energy density and pressure perturbations can be written in terms of the fields as
\cite{Hwang:2001fb, MW2008} 
\be 
\delta\rho=\sum_\alpha\Big(\dot{\vp}_\alpha\dot{\delta\vp}_\alpha-\dot{\vp}_\alpha^2\phi+V_{,\alpha}
\delta\vp_\alpha\Big)\,,
\ee
and
\be 
\delta P= \sum_\alpha\Big(\dot{\vp}_\alpha\dot{\delta\vp}_\alpha-\dot{\vp}_\alpha^2\phi - V_{,\alpha}
\delta\vp_\alpha\Big)\,,
\ee
and total adiabatic sound speed as
\begin{equation}
 c_\mathrm{s}^2 \equiv \frac{\dot{P}}{\dot{\rho}} = 1 + \frac{2 \sum_\alpha V_{,\alpha}
\dot{\vp}_\alpha}{ 3 H \sum_\beta \dot{\vp}_\beta^2} \,.
\end{equation}

Thus, we arrive at the following expression for the non-adiabatic pressure perturbation in the system consisting
of the two fields $\vp$ and $\chi$
\begin{align}
\label{eq:dPrel}
\delta P_{\rm nad} =
&-2(V_{,\vp}\delta\vp+V_{,\chi}\delta\chi)\\
&+\frac{2(V_{,\vp}\dot{\vp}+V_{,\chi}\dot{\chi})}{3H(\dot{\vp}^2+\dot{\chi}^2)}
\Big[(\dot{\vp}^2+\dot{\chi}^2)\phi-V_{,\vp}\delta\vp \nn \\
&-V_{,\chi}\delta\chi-\dot{\vp}\dot{\delta\vp}-\dot{\chi}\dot{\delta\chi}\Big]\,.\nn
\end{align}

Using one of the Einstein field equations we can relate the lapse function $\phi$ to field variables as
\be
H\phi=4\pi G(\dot{\vp}\delta\vp+\dot{\chi}\delta\chi)\,,
\ee
and so Eq.~(\ref{eq:dPrel}) becomes
%
\begin{align}
\delta P_{\rm nad} = &\frac{8\pi
G}{3H^2}(V_{,\vp}\dot{\vp}+V_{,\chi}\dot{\chi})(\dot{\vp}\delta\vp+\dot{\chi}\delta\chi)\\
&-2(V_{,\vp}\delta\vp+V_{,\chi}\delta\chi)\nn\\
&-\frac{2}{3H}\frac{(V_{,\vp}\dot{\vp}+V_{,\chi}\dot{\chi})}{(\dot{\vp}^2+\dot{\chi}^2)}
\Big[\dot{\vp}\dot{\delta\vp}+\dot{\chi}\dot{\delta\chi}+V_{,\vp}\delta\vp+V_{,\chi}\delta\chi\Big]\,.\nn
\end{align}

It is this definition of the non-adiabatic pressure perturbation that we calculate in the next section. In order
to compare with the comoving curvature perturbation ${\cal{R}}$, which is (in the flat gauge) 
\be 
{\cal{R}}=\frac{H}{\sum_\beta\dot{\vp}^2_\beta} \sum_\alpha \dot{\vp}_\alpha \delta\vp_\alpha \,,
\ee
we use a comoving entropy perturbation ${\cal{S}}$ introduced in Refs.~\cite{Gordon:2000hv, Malik2005}
and defined as 
\be 
{\cal{S}}=\frac{H}{\dot{P}}\delta P_{\rm nad}\,.
\ee

An alternative way to calculate perturbations in two-field inflationary models is to perform the field rotation into 
an adiabatic field, $\sigma$ and an isocurvature field $s$ as in Ref.~\cite{Gordon:2000hv}:
\begin{align}
\delta\sigma &= \cos\theta\delta\vp+\sin\theta\delta\chi\,,\\
\delta s &= -\sin\theta\delta\vp+\cos\theta\delta\chi\,,
\end{align}
where $\tan(\theta) = \dot{\chi}/\dot{\vp}$ and a comoving isocurvature perturbation can be then be
defined as
\be 
\widetilde{\mathcal{S}}=\frac{H}{\dot{\sigma}}\delta s\,,
\ee
where $\dot{\sigma}=\sqrt{\dot{\vp}^2 + \dot{\chi}^2}$,
The two isocurvature perturbation definitions ${\cal S}$ and $\tilde{\mathcal{S}}$ are then related
to one
another through \cite{Gordon:2000hv}
\footnote{Note that the final term in this expression differs slightly from that in Eq.\
(42) of Ref.~\cite{Gordon:2000hv}. In the slow roll limit when $\ddot{\sigma}$ is negligible the
expressions are equal.}
\begin{align}
\label{eq:Sevolution}
\cal S = &\frac{2}{3\dot{\sigma}^2(2V_{,\sigma} + 3H\dot{\sigma})}\times 
\Bigg\{ V_{,\sigma}\bigg[\dot{\sigma}\left(\dot{\delta\sigma}  
- \frac{8\pi G}{2H}\dot{\sigma}^2\delta\sigma\right) \nn \\
&-\ddot{\sigma}\delta\sigma\bigg]
- (3H\dot{\sigma}^2 \dot{\theta} + 2\dot{\sigma}V_{,\sigma} \dot{\theta}) \delta s \Bigg\}\,,
\end{align}
where $V_{,\sigma} = \sum_\alpha V_{,\alpha}\dot{\vp}_\alpha /\dot{\sigma}$ and we have substituted
for $H\phi = 4\pi G \dot{\sigma}\delta \sigma$.
In Section~\ref{sec:results} results are given for both $\mathcal{S}$ and $\widetilde{\mathcal{S}}$
to show the differences in evolution. In the slow roll and large scale limits
$\mathcal{S}$ and $\widetilde{\mathcal{S}}$ become equivalent and $\delta s$ is
an isocurvature direction. In general this is not the case as can be seen in
Section~\ref{sec:results} when slow roll breaks down at the end of inflation.
It is clear from \eq{eq:Sevolution} that even when the
isocurvature perturbation $\delta s$ is highly suppressed, $\cal{S}$ can still evolve due to the
dependence on the adiabatic component $\delta \sigma$. This is an important distinction between the
two definitions of isocurvature which is not often recognized. 

\section{Numerical procedure}
\label{sec:numerical}
We have used the {\sc Pyflation} numerical package to obtain the results in this paper
\cite{pyflation,hustonmalik2}. This package uses the Numerical and Scientific Python libraries
\cite{numpy,oliphant:10,scipy} and is based around a Runge-Kutta ODE solver
\cite{abramowitz+stegun}. The {\sc Pyflation} package has been updated to solve first order
systems of multiple inflationary fields and this new version is now available for download under an
open source license \cite{pyflation}.

As described above we evolve
the perturbed field values $\delta\vp$ and $\delta \chi$ in contrast with other approaches. In
fact the quantum nature of the perturbations on sub-horizon scales means that for a multi-field
system the operators for each of the $N$ fields contain $N$ independent annihilation
operators as 
\begin{equation}
 \widehat{\delta\vp}_\alpha = \sum_\beta \xi_{\alpha \beta} \hat{a}_\beta \,.
\end{equation}
Our numerical system solves the evolution equation for the $N\times N$ mode function matrix
$\xi_{\alpha \beta}$ following Ref.~\cite{Salopek:1988qh}. 
It is common to replicate this behaviour without employing the full mode matrix by running the
simulation multiple times, with each of the fields in turn set to zero initially. If the matrix
approach or the multiple runs approach are not used the numerical results will be incomplete as
they will not account correctly for the cross-terms in the mode matrix (e.g. $\xi_{\phi \chi}$ in
the two-field case) (see Refs.~\cite{Lalak:2007vi,Avgoustidis:2011em} and the notes contained there
on previous numerical calculations).

The Klein-Gordon equations for the scalar field perturbations must be rewritten in terms of the mode
function matrices \cite{Salopek:1988qh}. For clarity these equations are presented here using
coordinate time, however 
in the numerical system the time variable is the number of efolds $\mathcal{N}$, and all the ODEs
are expressed in terms of derivatives with respect to $\mathcal{N}$. For the first order mode
function matrix elements the Klein-Gordon equation is
\begin{align}
 \ddot{\xi}_{\alpha \beta} &+ 3 H \dot{\xi}_{\alpha \beta} + \left(\frac{k}{a}\right)^2 \xi_{\alpha
\beta} + \sum_\gamma \Bigg\{ V_{,\alpha \gamma} \nonumber \\
&+ \frac{8\pi G}{H}\left(\dot{\vp}_{0\alpha}
V_{,\gamma} + \dot{\vp}_{0\gamma} V_{,\alpha} + \frac{8\pi G}{H} \dot{\vp}_{0\alpha} 
\dot{\vp}_{0\gamma} V \right) \Bigg\} \xi_{\gamma \beta} \nonumber \\
&= 0\,.
\end{align}

The initial quantum state is
taken to be the Bunch-Davies vacuum state with the scalar fields uncorrelated
\begin{equation}
 \xi_{\alpha \beta} = \frac{\sqrt{8\pi G}}{a\sqrt{2k}} e^{-i k \eta} \delta_{\alpha \beta}\,,
\end{equation}
where $\eta=\int dt/a$ is conformal time.
All the quantities calculated from the field perturbations inherit this mode matrix structure so
for example $\delta P_\mathrm{nad} = \sum_\alpha \delta P_\mathrm{nad\,\alpha} \hat{a}_\alpha$. 

When using the mode matrix approach it is important to calculate the power spectra of quantities
correctly using the commutation relations of the quantum operators in order to account for the
cross-terms. For example the power spectrum of comoving curvature perturbations
$\mathcal{P}_\mathcal{R}$ is given by
\begin{equation}
 \mathcal{P}_\mathcal{R}(k) = \frac{2\pi^2}{k^3} \frac{H^2}{\left(\sum_\alpha \dot{\vp}_\alpha^2
\right)^2} \sum_{\beta,\gamma} \left\{ \dot{\vp}_\beta \dot{\vp}_\gamma 
\sum_\lambda \xi_{\beta \lambda}^{} \xi_{\gamma \lambda}^* \right\}\,.
\end{equation}

\section{Results}
\label{sec:results}

In this section we present our numerical results on three inflationary potentials involving two
scalar fields. These potentials have been well studied in the literature and we have used these
previous results to ensure our code is working correctly. 

\subsection{Double quadratic inflation}

\begin{figure}
 \centering
 \includegraphics[width=0.5\textwidth]{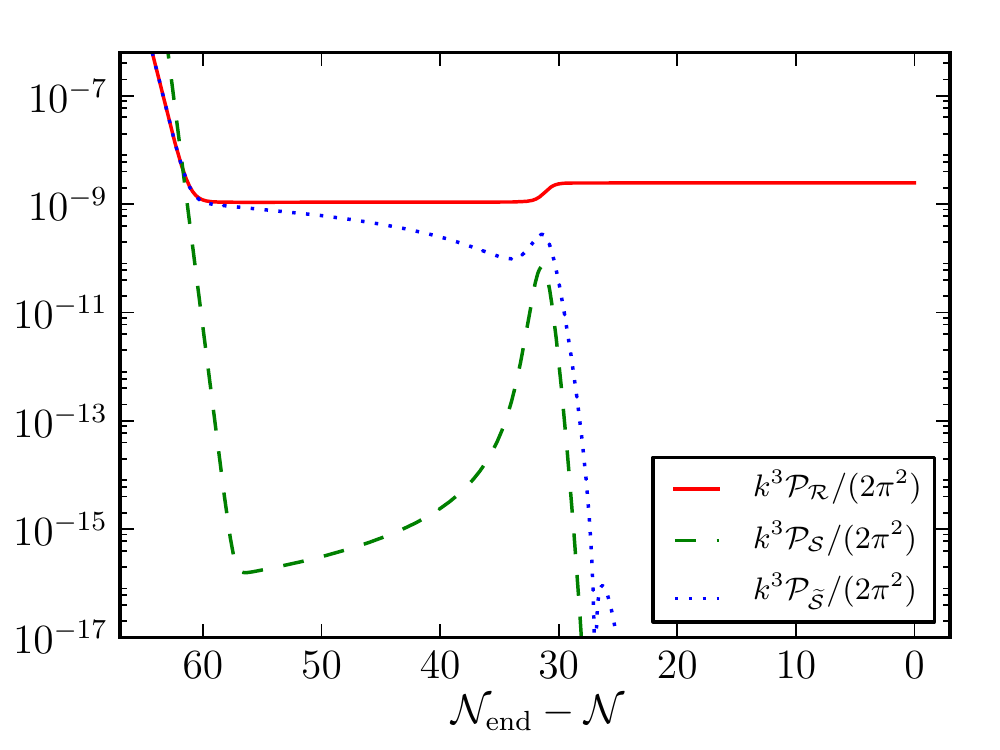}
 \caption{A comparison of the power spectra 
$\mathcal{P}_\mathcal{R}$ (red straight line), $\mathcal{P}_\mathcal{S}$ (green dashed line) and
$\mathcal{P}_{\tilde{\mathcal{S}}}$ (blue dotted line) for the double quadratic
potential at the {\sc Wmap} pivot scale.}
 \label{fig:quad-PrSSalt_N}
\end{figure}

\begin{figure}
 \centering
 \includegraphics[width=0.5\textwidth]{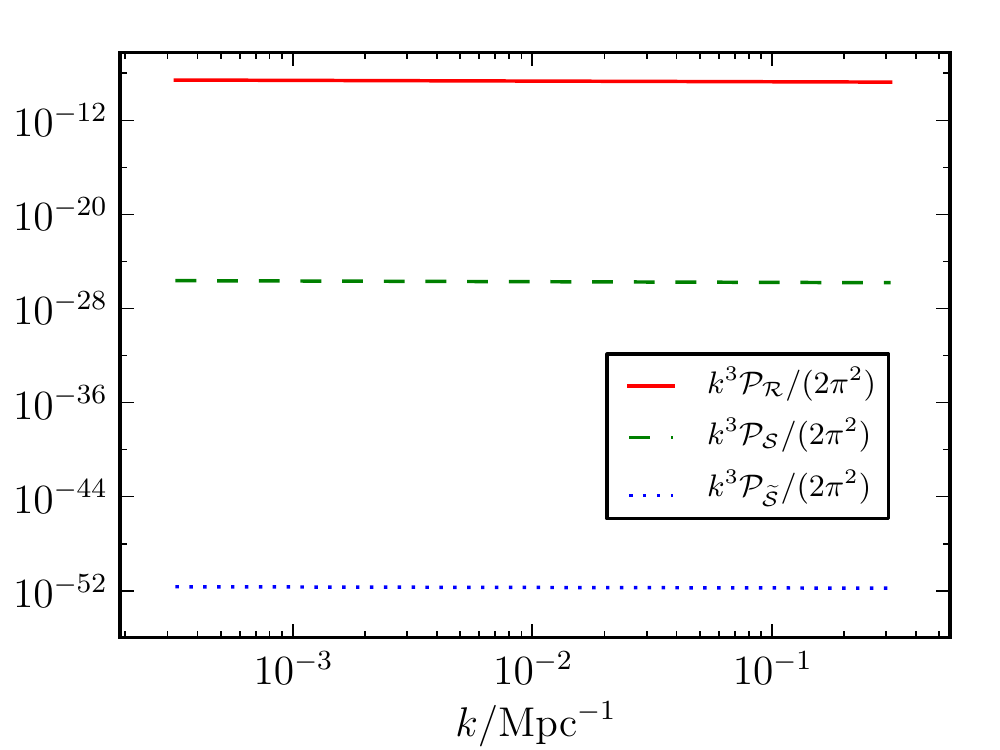}
 \caption{A comparison of the power spectra 
$\mathcal{P}_\mathcal{R}$ (red straight line), $\mathcal{P}_\mathcal{S}$ (green dashed line) and
$\mathcal{P}_{\tilde{\mathcal{S}}}$ (blue dotted line) for the double quadratic
potential in terms of $k$ at the end of inflation.}
 \label{fig:quad-PrSSalt_kend}
\end{figure}

\begin{figure}
 \centering
 \includegraphics[width=0.5\textwidth]{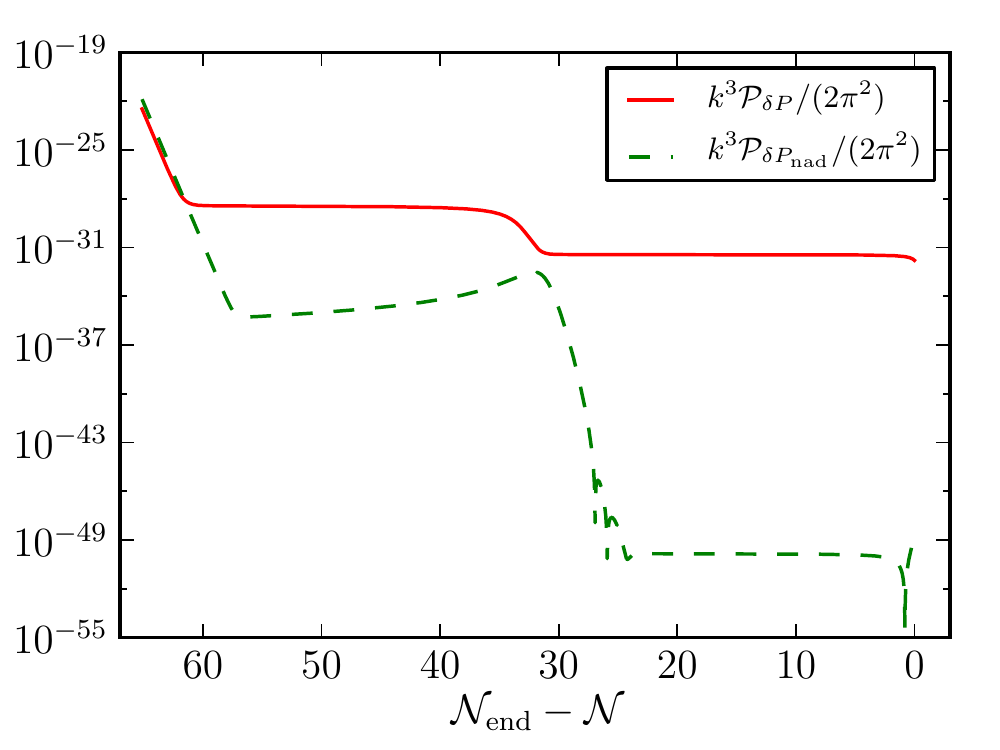}
 \caption{A comparison of the power spectra 
of $\delta P$ (red straight line) and $\delta P_\mathrm{nad}$ (green dashed line)
for the double quadratic
potential at the {\sc Wmap} pivot scale.}
 \label{fig:quad-dPnaddP_N}
\end{figure}

We first consider the well-studied case of double quadratic inflation \cite{Langlois:1999dw} for
which the potential takes the form
\be 
V(\vp,\chi)=\frac{1}{2}m_\vp^2\vp^2+\frac{1}{2}m_\chi^2\chi^2\,.
\ee
We consider the particular case when $m_\chi = 7 m_\vp$ as studied in
Refs.~\cite{Avgoustidis:2011em, Lalak:2007vi}, and we set $m_\vp=1.395\times 10^{-6} \Mpl$
in order for the curvature power spectrum at the pivot scale to match the WMAP7 maximum likelihood
value of $2.45\times 10^{-9}$. For the results shown here in
Figures~\ref{fig:quad-PrSSalt_N},\ref{fig:quad-PrSSalt_kend} and \ref{fig:quad-dPnaddP_N} the
initial field values are $\vp_0 = \chi_0 = 12 \Mpl$ and the derivatives of the fields are
given their slow roll values. The spectral index of the curvature perturbations for these parameter
choices is approximately $n_\mathcal{R} \simeq 0.937$ (with no running allowed).

In this model the inflationary dynamics are originally dominated by the $\chi$ field until around
$30$ efolds before the end of inflation when the $\vp$ field becomes dominant. This is associated
with a rise in the curvature perturbation amplitude as shown in Figure~\ref{fig:quad-PrSSalt_N}
where the power spectra for the comoving curvature perturbation and the two definitions of
isocurvature perturbations are compared. In all the graphs in this section results are plotted
against either the number of efoldings $\mathcal{N}$ left until the end of inflation
($\mathcal{N}$) or the comoving wavenumber $k$ of the perturbation modes in units of inverse
Megaparsecs. When plotted against $\mathcal{N}$ the mode shown is the WMAP pivot scale $k=0.002
\Mpc^{-1}$.
 
The evolution of $\widetilde{\mathcal{S}}$ matches
that found in Ref.~\cite{Avgoustidis:2011em} with the amplitude dropping off significantly after
the change-over. This is a good consistency check for our mode function matrix approach as that
paper used the alternative multiple-run method as described in Section~\ref{sec:numerical} and
performed the field redefinition into $\sigma$ and $s$ before the numerical run. 

The evolution of $\mathcal{S}$ is not similar to $\widetilde{\mathcal{S}}$ especially as the mode
crosses the horizon around
$60$ efoldings before the end of inflation. The amplitude of $\mathcal{S}$ continues to reduce for a
few
efoldings after horizon crossing before rapidly increasing to reach a peak at the cross-over
time. It then rapidly drops again and at the end of inflation is many orders of magnitude smaller
than $\mathcal{R}$ as shown in Figure~\ref{fig:quad-PrSSalt_kend}. The amplitude of
$\widetilde{\mathcal{S}}$ is
even further suppressed in comparison but the numerical value at such a relatively small level must
be treated with caution. We have also calculated the spectral index of the isocurvature
perturbations using a similar expression to $n_\mathcal{R}$:
\begin{equation}
 n_\mathcal{S} = \frac{d\log(\mathcal{P}_\mathcal{S})}{d\log(k)} + 1\,.
\end{equation}
As one might guess from Figure~\ref{fig:quad-PrSSalt_kend} the two spectral indices have very
similar values. This might be expected given the dependence of $\delta P_\mathrm{nad}$
and therefore $\mathcal{S}$ on the combinations of the fields which are also present in
$\mathcal{R}$, but this result will be very useful for future studies of the generation of
vorticity as discussed further in Section~\ref{sec:discussion}.

The quantity of interest we described in Section~\ref{sec:intro} is the non-adiabatic pressure
perturbation $\delta P_\mathrm{nad}$ and this is shown in Figure~\ref{fig:quad-dPnaddP_N} along with
the total pressure perturbation $\delta P$. The evolution of $\delta P_\mathrm{nad}$ is broadly
similar to $\mathcal{S}$, after a sharp fall on crossing the horizon the amplitude increases until
the
cross-over point at which time it sharply falls again. This is to be expected as the rate of change
in $\mathcal{R}$ is proportional to $\delta P_\mathrm{nad}$ \cite{MW2008,Malik2005}.

\subsection{Double quartic inflation}

\begin{figure}
 \centering
 \includegraphics[width=0.5\textwidth]{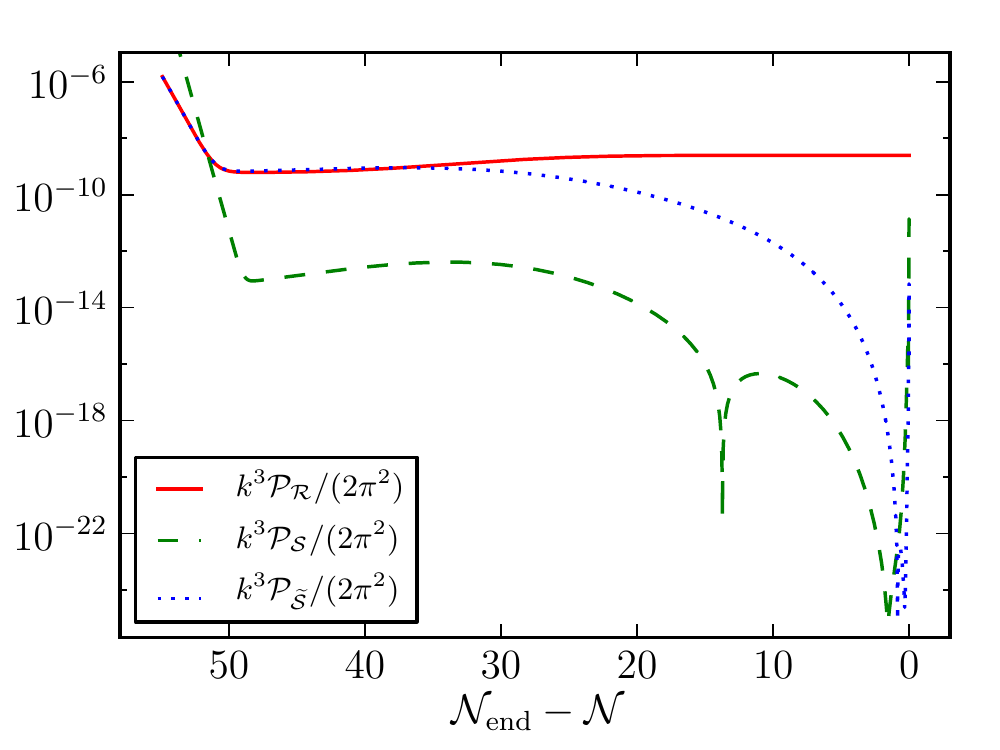}
 \caption{A comparison of the power spectra 
$\mathcal{P}_\mathcal{R}$ (red straight line), $\mathcal{P}_\mathcal{S}$ (green dashed line) and
$\mathcal{P}_{\tilde{\mathcal{S}}}$ (blue dotted line) for the double quartic
potential at the {\sc Wmap} pivot scale.}
 \label{fig:quart-PrSSalt_N}
\end{figure}

\begin{figure}
 \centering
 \includegraphics[width=0.5\textwidth]{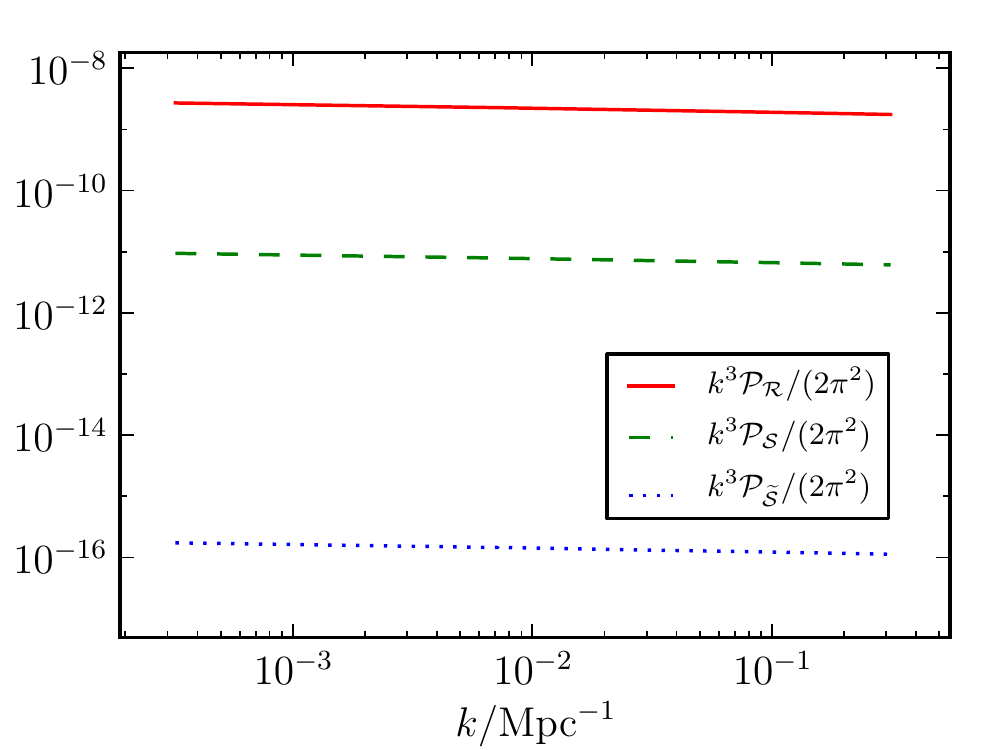}
 \caption{A comparison of the power spectra 
$\mathcal{P}_\mathcal{R}$ (red straight line), $\mathcal{P}_\mathcal{S}$ (green dashed line) and
$\mathcal{P}_{\tilde{\mathcal{S}}}$ (blue dotted line) for the double quartic
potential in terms of $k$ at the end of inflation.}
 \label{fig:quart-PrSSalt_kend}
\end{figure}

\begin{figure}
 \centering
 \includegraphics[width=0.5\textwidth]{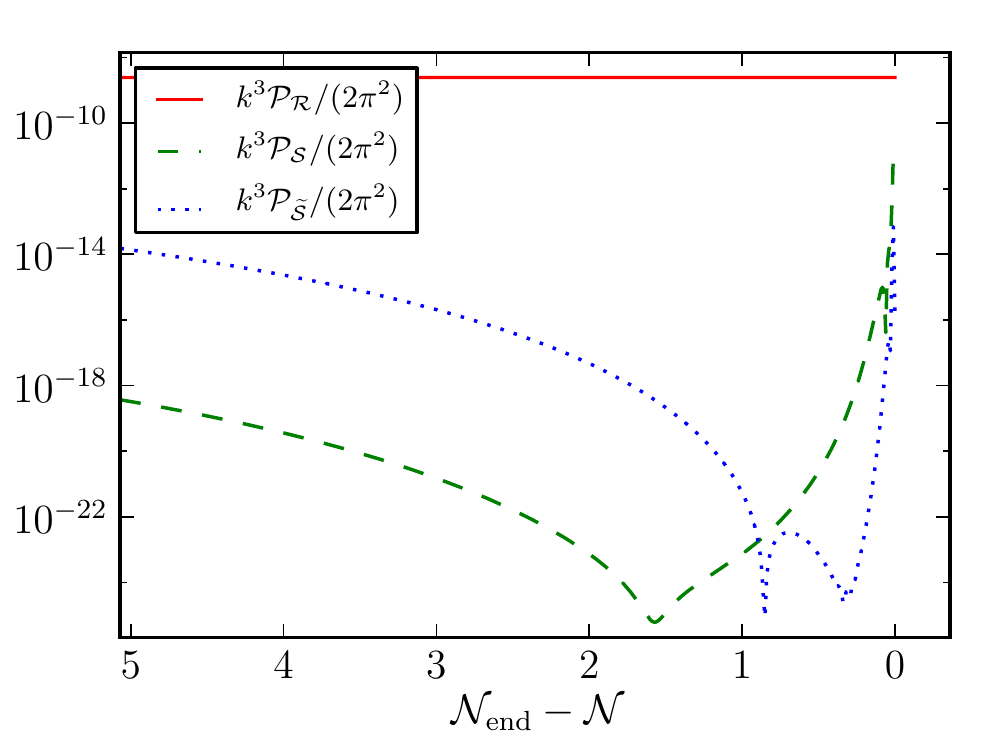}
 \caption{A comparison of the power spectra 
$\mathcal{P}_\mathcal{R}$ (red straight line), $\mathcal{P}_\mathcal{S}$ (green dashed line) and
$\mathcal{P}_{\tilde{\mathcal{S}}}$ (blue dotted line) for the double quartic
potential at the {\sc Wmap} pivot scale over the last 5 e-foldings of inflation.}
 \label{fig:quart-PrSSalt_N_5efolds}
\end{figure}

\begin{figure}
 \centering
 \includegraphics[width=0.5\textwidth]{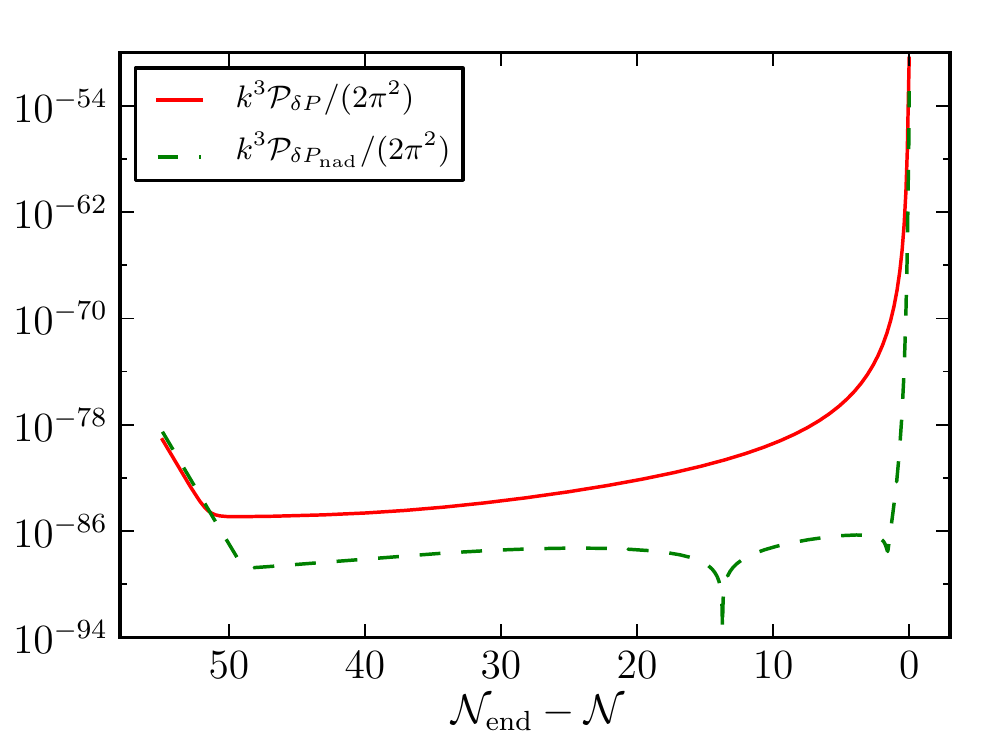}
 \caption{A comparison of the power spectra 
of $\delta P$ (red straight line) and $\delta P_\mathrm{nad}$ (green dashed line)
for the double quartic
potential at the {\sc Wmap} pivot scale.}
 \label{fig:quart-dPnaddP_N}
\end{figure}

Next, we study a special case of hybrid inflation that was considered in Refs.~\cite{Avgoustidis:2011em, Kodama:2011vs}.
The potential for this model takes the form
\be 
V(\vp,\chi)=\Lambda^4\Bigg[
\Bigg(1-\frac{\chi^2}{v^2}\Bigg)^2+\frac{\vp^2}{\mu^2}+\frac{2\vp^2\chi^2}{\vp_{\rm c}^2v^2}\Bigg]\,,
\ee
with the parameter values $v=0.10 \Mpl$, $\vp_{\rm c}=0.01 \Mpl$ and $\mu=10^3 \Mpl$. In
this case $\Lambda$ can be normalised to match the WMAP results by setting $\Lambda=2.36\times
10^{-4} \Mpl$. The fields are started at the initial values $\vp_0=0.01 \Mpl$ and $\chi_0=1.63\times
10^{-9} \Mpl$. For these choices the spectral index at the end of inflation is around
$n_\mathcal{R}=0.932$ when no running is allowed.

As shown in Figure~\ref{fig:quart-PrSSalt_N} the behaviour in this case is markedly different.
$\mathcal{R}$ continues to evolve outside the horizon and $\widetilde{\mathcal{S}}$ has a greater
amplitude than
$\mathcal{R}$ for a few efoldings after horizon crossing, again matching the results of
Ref.~\cite{Avgoustidis:2011em}. The
evolution of $\mathcal{S}$ outside the horizon is unremarkable until around the final five efoldings
of
inflation when a sharp increase in amplitude occurs. Close inspection of these final efoldings in
Figure~\ref{fig:quart-PrSSalt_N_5efolds} shows that the amplitude of $\mathcal{S}$ rebounds to reach
significant levels at the end of inflation. From Figure~\ref{fig:quart-PrSSalt_kend} we can see
that $\mathcal{P}_\mathcal{S}$ is of the order of a few percent of $\mathcal{P}_\mathcal{R}$. 

The evolution of the pressure perturbations is also very different from the standard double
quadratic case. Figure~\ref{fig:quart-dPnaddP_N} shows $\delta P$ increasing throughout the
super-horizon evolution while the level of $\delta P_\mathrm{nad}$ is relatively constant until the
final few efoldings.

\subsection{Product exponential}

\begin{figure}
 \centering
 \includegraphics[width=0.5\textwidth]{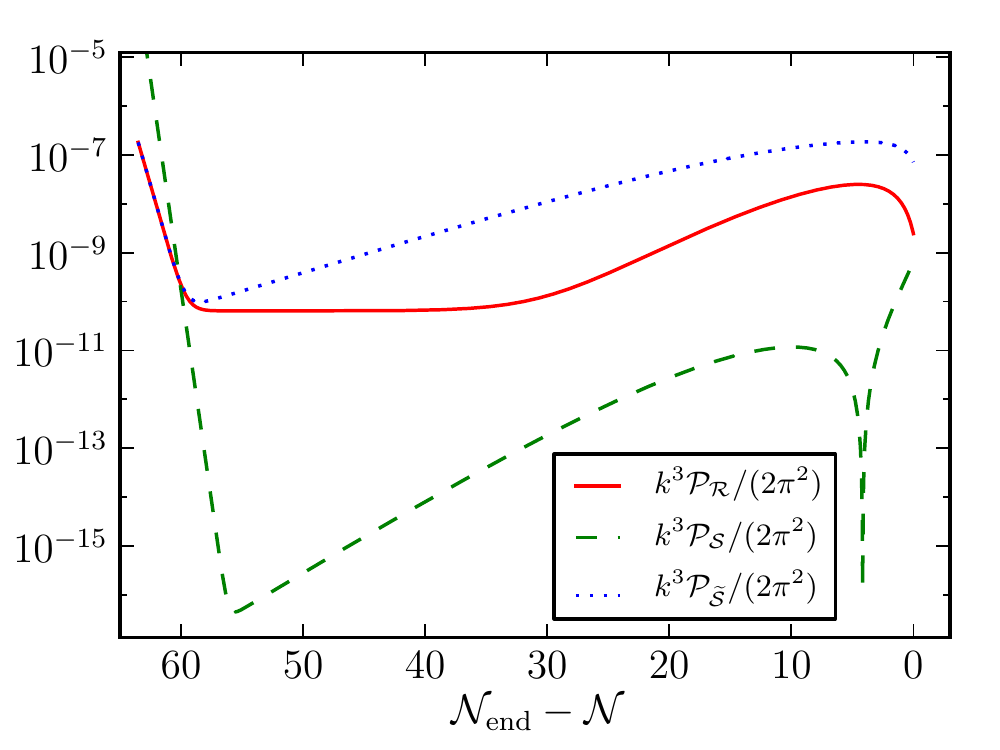}
 \caption{A comparison of the power spectra 
$\mathcal{P}_\mathcal{R}$ (red straight line), $\mathcal{P}_\mathcal{S}$ (green dashed line) and
$\mathcal{P}_{\tilde{\mathcal{S}}}$ (blue dotted line) for the product exponential
potential at the {\sc Wmap} pivot scale.}
 \label{fig:prodexp-PrSSalt_N}
\end{figure}

\begin{figure}
 \centering
 \includegraphics[width=0.5\textwidth]{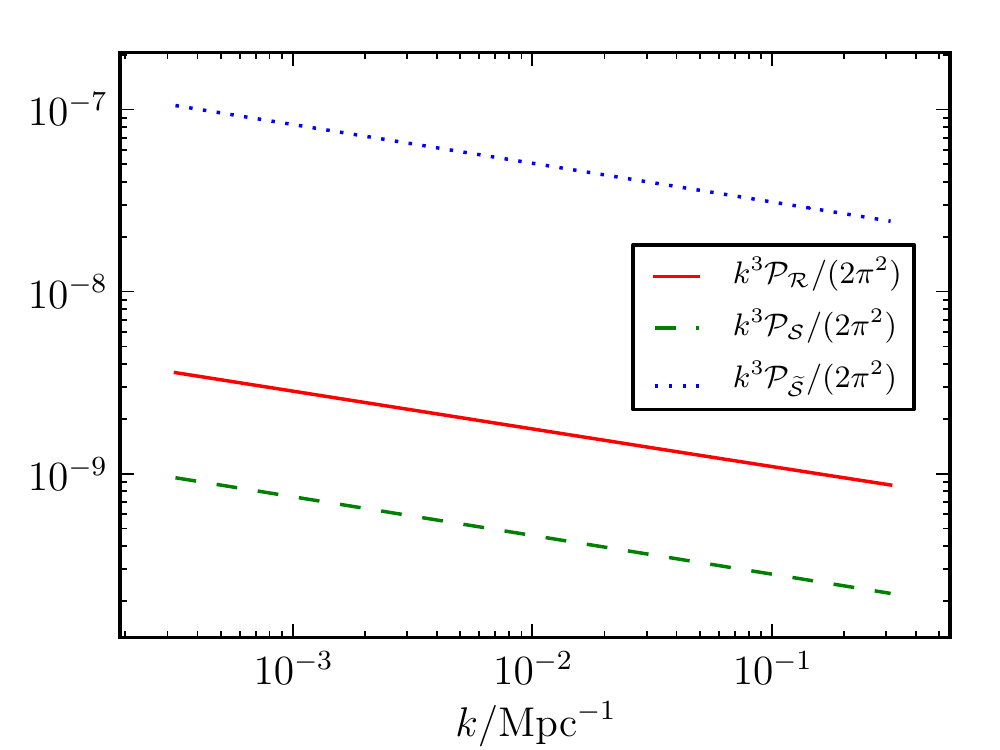}
 \caption{A comparison of the power spectra 
$\mathcal{P}_\mathcal{R}$ (red straight line), $\mathcal{P}_\mathcal{S}$ (green dashed line) and
$\mathcal{P}_{\tilde{\mathcal{S}}}$ (blue dotted line) for the product exponential
potential in terms of $k$ at the end of inflation.}
 \label{fig:prodexp-PrSSalt_kend}
\end{figure}

\begin{figure}
 \centering
 \includegraphics[width=0.5\textwidth]{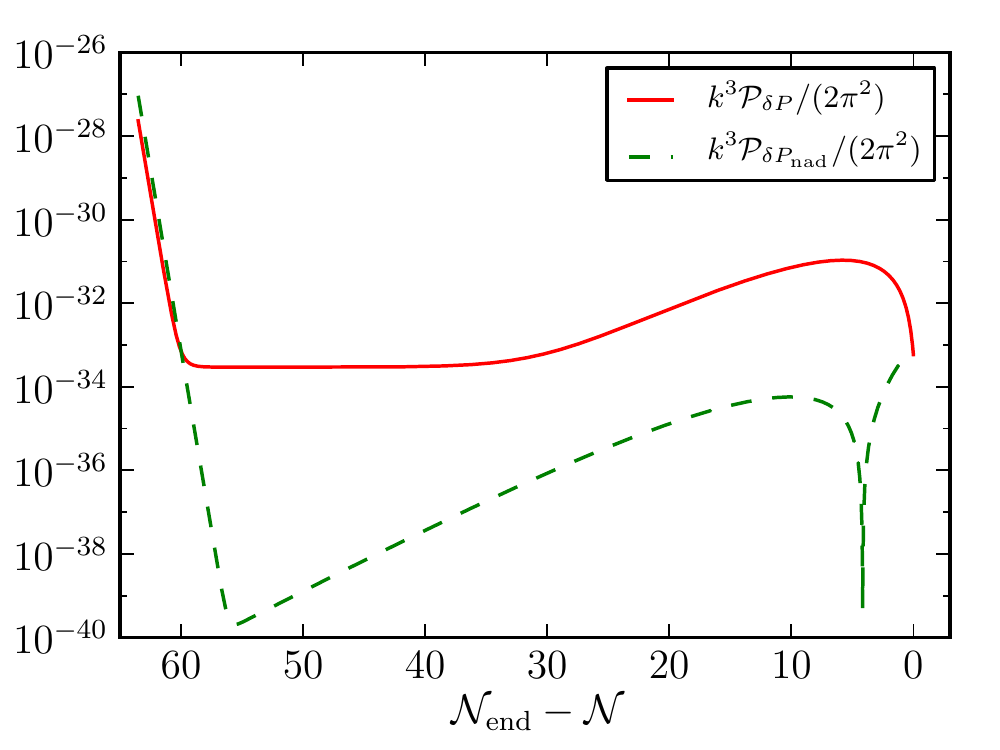}
 \caption{A comparison of the power spectra 
of $\delta P$ (red straight line) and $\delta P_\mathrm{nad}$ (green dashed line)
for the product exponential
potential at the {\sc Wmap} pivot scale.}
 \label{fig:prodexp-dPnaddP_N}
\end{figure}

Finally we investigate a quite particular potential which has appeared recently in the literature
in connection with the production of non-Gaussianity \cite{Byrnes:2008wi,Elliston:2011dr}.
This is a two field model with the product separable form
\begin{equation}
 V = V_0 \vp^2 e^{-\lambda \chi^2}\,,
\end{equation}
where we have set $\lambda = 0.05 /\Mpl^2$ and normalized the results to the WMAP value by setting
$V_0 =5.37\times 10^{-13} \Mpl^2$. For the results shown here the initial field values are $\vp_0=18
\Mpl$ and $\chi_0=0.001 \Mpl$.

This model has been used as an example of a product separable potential in which it is possible to
generate a large negative amount of non-Gaussianity\footnote{The sign convention used is that of
the WMAP team.}. Ref.~\cite{Byrnes:2008wi} followed the evolution using the $\delta\mathcal{N}$
formalism until the slow roll approximation becomes invalid before the end of inflation and showed
that $f_\mathrm{NL}$ values of the order of $-35$ can be obtained during
inflation. In Ref.~\cite{Elliston:2011dr} the authors showed that
extremely large oscillations of the value of $f_\mathrm{NL}$ occur after slow
roll breaks down and on into the reheating phase. Because of this and the
presence of isocurvature it is not clear whether the values of $f_\mathrm{NL}$
during inflation can be related to the observed values.
In Figure~\ref{fig:prodexp-PrSSalt_N} the late time evolution of $\mathcal{R}$
can be clearly seen
as inflation ends. What is also very clear from our results is that the isocurvature perturbations
measured by $\widetilde{\mathcal{S}}$ have significantly larger amplitudes than the adiabatic
perturbations. The
magnitude of $\mathcal{S}$ is smaller than $\mathcal{R}$ but it is a sizeable fraction at the end of
inflation as shown in Figure~\ref{fig:prodexp-PrSSalt_kend} and puts this model in conflict with
the restrictions on isocurvature fraction from the CMB.

It is also clear from Figure~\ref{fig:prodexp-PrSSalt_kend} that the scale dependence of the
curvature perturbations is not close to the maximum likelihood value from the WMAP 7 year data
release. At the end of inflation the spectral index of the curvature perturbations is around
$n_\mathcal{R}=0.794$ when no running is allowed\footnote{$n_\mathcal{S}$ is again very similar to
$n_\mathcal{R}$ in this case.}. However, the spectral index changes considerably in the last few
efoldings of inflation. The effects of reheating can not be neglected for this model and it is
possible that after reheating the value of $n_\mathcal{R}$ will be at least closer to the current
observational limits. This is highlighted in more detail in
Ref.~\cite{Dias:2011xy}.

As is now familiar the amplitude of $\delta P_\mathrm{nad}$ drops as the mode crosses the horizon
and $\mathcal{R}$ stops decreasing. The amplitude of $\delta P_\mathrm{nad}$ then rises
through the super-horizon evolution before crossing through zero as the amplitude of $\mathcal{R}$
reaches its maximum, and finally reaching levels comparable with $\delta P$.

In this section we have shown the results of our numerical simulations for three different
two field potentials. We have seen that the evolution of $\mathcal{S}$ and $\delta P_\mathrm{nad}$
is markedly different to that of $\widetilde{\mathcal{S}}$ and $\delta s$. We have focussed for the
most part on models in which a significant fraction of isocurvature is present near the end of
inflation and these models will provide a good starting point for the generation of vorticity in
the post-inflationary era. We have also seen the third model, the product exponential potential, is
in some tension with the observational limits on isocurvature fraction and also the spectral index.
The effects of reheating in this model are beyond the scope of this work but there are suggestions,
especially from the results of Ref.~\cite{Elliston:2011dr} that the changes during this phase would
be worthy of further consideration.

 \section{Discussion}
 \label{sec:discussion}
 
In this paper we have studied the spectra of entropy, or isocurvature, perturbations in various
inflationary models,
focussing on models consisting of two scalar fields. We have undertaken a fully numerical study
without resorting to an expansion in slow roll parameters. We have shown that the non-adiabatic
pressure perturbation $\delta P_\mathrm{nad}$ and its related quantity $\cal{S}$ can evolve in a
quite different way to the isocurvature mode $\delta s$ which is frequently used.

The study of isocurvature perturbations has been very popular recently. Other work includes looking at isocurvature
perturbations from multi-field inflation with non-canonical kinetic terms \cite{Lalak:2007vi,
Gao:2009qy, Peterson:2010np}, or using the 
field redefinition, first considered in Ref.~\cite{Gordon:2000hv}, into adiabatic and 
entropic modes \cite{Tsujikawa:2002qx, Avgoustidis:2011em}.

We have studied three inflationary potentials in this paper and described the evolution of the
isocurvature perturbations until the end of inflation.
After inflation, of course, the universe will need to go through a phase of reheating during
which the scalar fields
driving inflation are converted into the standard model particles. This phase will have an effect 
on the spectrum of the non-adiabatic perturbations that we have obtained during inflation, and could
cause the resonant growth of the entropy perturbations (see, e.g., Ref.~\cite{Allahverdi:2010xz}).
However, since there is no agreed upon mechanism of reheating, we do not study that here, instead
focussing on determining which inflationary models in their own right
can generate sizeable entropy perturbations. 

As mentioned in the Introduction, one of the motivations for undertaking this study is the recent
work highlighting the importance of entropy
perturbations on the generation of vorticity at second order in cosmological perturbation theory. As
shown in detail in   Ref.~\cite{vorticity}, second order vorticity, $\w_{2ij}$ evolves (during
radiation domination) according to
\be 
\dot{\w}_{2ij}-H\w_{2ij}\propto\delta\rho_{,[j}\delta P_{{{\rm nad}}1,i]}\,,
\ee
where the subscripts denote the order in perturbation theory and the square brackets denote anti-symmetrisation.
So, vorticity is sourced by the coupling between linear energy density and entropy gradients. The perturbation
to the energy density is well-known, and forms the basis of much cosmological study. However, entropy perturbations
are rarely studied in cosmology in this context and so the results presented here will allow us to advance this field
of study. 

At least for the models studied in this paper, the scale dependence of $\delta P_\mathrm{nad}$ is
very similar to that of the adiabatic curvature perturbations. This could have been anticipated but
up to now this relationship has not been assumed in the literature on vorticity generation and
instead various ans\"{a}tze have been made for how $\delta P_\mathrm{nad}$ varies with $k$. We hope
that the results in this paper will be able to inform future work on the generation of vorticity and
other non-linear effects using non-adiabatic pressure perturbations.

As we have seen, for some models it is expected that reheating could strongly affect both the
adiabatic and isocurvature results. The next step is to implement a straightforward reheating
method, with the inflaton decaying into a small number of species. It has also been shown that
non-adiabatic pressure can be generated during a radiation phase even when the initial conditions
are purely adiabatic \cite{Brown:2011dn}. By adding in the results presented here for isocurvature
from the end of inflation it will be possible to refine these simulations and allow better
comparisons with the observational data at later times in the universe's evolution.

In conclusion we have shown the evolution of two different types of isocurvature perturbation for a
range of two field models of inflation. Our numerical results indicate that the evolution of $\delta
P_\mathrm{nad}$ can be quite different from the isocurvature mode $\delta s$ which is usually
considered in the literature. Our results are significant when considered in conjunction with the
generation of vorticity by $\delta P_\mathrm{nad}$, and the effect this may have on the
interpretation of observational results from the CMB, in particular B-mode polarisation.  
\\
\bigskip

\section*{Acknowledgements}
The authors are grateful to Karim Malik, David Mulryne, Paul Saffin, David
Seery and Lixin Xu for useful discussions and for comments on
a previous version of this manuscript. 
IH is supported by the STFC under Grant
ST/G002150/1, and AJC is funded by the Sir Norman Lockyer
Fellowship of the Royal Astronomical Society.

\bibliography{isocurvpapers}

\end{document}